\documentclass[a4paper,aps,preprintnumbers,twocolumn,amsmath,amssymb,pra,showpacs]{revtex4}
\usepackage{bbm}
\usepackage{mathrsfs}
\usepackage{amsmath}
\usepackage[dvips]{graphicx}
\usepackage{epsfig}
\usepackage[dvips]{graphics,graphicx}

\newcommand{\be}{\begin{equation}}
\newcommand{\ee}{\end{equation}}
\newcommand{\bea}{\begin{eqnarray}}
\newcommand{\eea}{\end{eqnarray}}
\newcommand{\bes}{\begin{split}}
\newcommand{\ees}{\end{split}}

\renewcommand{\vec}[1]{\mathbf{#1}}

\newcommand{\tr}{\operatorname{Tr}}

\usepackage[dvips]{graphics,graphicx}

\begin{document}
\title{Phase Diagrams for Spin-1 Bosons in an Optical Lattice}
\author{Ming-Chiang Chung and Sungkit Yip}
\affiliation{Institute of Physics, Academia Sinica, Taipei 11529, Taiwan}

\begin{abstract}
  In this paper, the phase diagrams of a polar spin-1 Bose gas in a three-dimensional optical lattice with   linear and quadratic Zeeman effects both at zero and  finite temperatures are obtained within mean-field theory. The phase diagrams can be regrouped to two different parameter regimes depending on the  magnitude of  the quadratic Zeeman effect $Q$.  For large $Q$, only a first-order phase transition from the nematic  (NM) phase to the fully magnetic (FM) phase  is found, while in the case of small $Q$, a first-order phase transition from the nematic phase to the partially magnetic (PM) phase , plus a second-order phase transition from the PM phase to the FM phase is obtained. If a net magnetization in the system exists, the first-order phase transition causes a coexistence of two phases and phase separation: for large $Q$, NM and FM phases and for small $Q$, NM  and PM phases. The phase diagrams in terms of net magnetization are also obtained. 
\end{abstract}

\date{\today}

\maketitle

\section{Introduction}

The study of cold atoms in optical lattices has captured a lot of
recent attention.  A primary motivation is to study the strongly repulsive
(two spin species) Fermi Hubbard model in the regime of close to one
atom per lattice site in two dimension, a system which is believed by many to capture
the most essential physics of the high temperature oxide
superconductors \cite{Lee06}.  Much progress has already been made
towards this goal, in particular the Mott insulating phase in three dimension has
already been obtained \cite{Jordens08,Schneider08}.  However, the
expected anti-ferromagnetic Neel ordering has not yet been reported,
perhaps due to the difficulty in cooling fermions.

On the other hand, there are also substantial interests in studying
Bosons with spins in the Mott insulating regime in an optical
lattice.  There have already been quite a number of experimental
studies on spinor Bose-Einstein condensates (without optical
lattice) \cite{MIT,Spin1a,Chang,Sengstock04,Kuwamoto04,Griesmaier05}. Mott
insulating state of Bosons with frozen spin degree of freedom has
also been achieved experimentally.\cite{Greiner02}  Hence, one can
be hopeful that we can study experimentally Bosons with spin in an
optical lattice in the Mott regime, where though there is no net
mass transport possible, the spin degree of freedom is still active.
Due to the finite tunneling amplitude and hence exchange interaction
between bosons on neighboring sites, one again expect the
possibility of studying quantum magnetism and ordering in these
systems.  Moreover, it can easily be seen that the spin Hamiltonian
realized in these systems would be very different from their
counterpart in solid state magnetic systems.  For example, for
spin-1 atoms, the Hamiltonian coupling neighboring spins ${\bf
S}_{i,j}$ is of the form \cite{Yip03,Imambekov03} $  J ( {\bf S}_i
\cdot {\bf S}_j )
    \  +  \ K ( {\bf S}_i \cdot {\bf S}_j )^2 $
with $K$ of the same order as $J$.  This is very different from the
usual Heisenberg Hamiltonian $  J ( {\bf S}_i \cdot {\bf S}_j ) $
which well describes  electronic spin interaction in solids.
Indeed, a large number of theoretical papers have already been
devoted to the subject of the spin physics in these systems. (see
\cite{Spinhalf,Spin1,Rizzi05,Harada07,Spin2,Spin3} and references
therein).

In this paper, we consider spin-1 Bosons in an isotropic
three-dimensional optical lattice in the Mott regime of one particle
per site.  We are in particular interested in the case of
anti-ferromagnetic interaction between the atoms, as in the case 
$^{23}Na$.  The Hamiltonian \cite{Yip03,Imambekov03} correspond to
$J < 0$, $K <0$ with $|J| < |K|$.  This spin Hamiltonian has already
been considered in the literature even before the field of cold
atoms \cite{ChenLevy,old}.  A general consensus was that, at low
temperatures, the system would order in a nematic state which breaks
rotational symmetry but has no net spin on any site. (The dimer
state, the ground state in one-dimension \cite{Yip03,Rizzi05}, is
unstable towards the nematic state with sufficiently strong coupling
between neighboring chains \cite{Harada07}). However, there are some
issues in cold-atom systems which were not considered in these
works, and we would like to remedy a few of these in this paper. One
is the existence of finite magnetic fields in realistic experiments.
This magnetic field produces a "quadratic Zeeman"
effect\cite{linear}, which lifts the energy degeneracy between two
atoms in the $m_f = 0$ hyperfine sublevel versus one each in $m_f =
\pm 1$.  The other consideration is that, in the time scale of the
experiment, the net "magnetization", namely the sum of $m_f$ over
all the particles, is conserved.  This "constant magnetization"
constraint was usually ignored in previous studies. Since in
particular the nematic state itself carries no magnetization, it is
natural to ask what is the thermodynamical state of the system if
one is constrained to have a finite net magnetization.  Besides
intrinsic interest, this issue may be relevant since a realistic
experiment may not always have exactly equal numbers of $m_f = \pm
1$ atoms in its initial preparation. Lastly, one need to consider
finite temperatures.  The nematic state can now tolerate some net
magnetization via thermally excited particles, and it is of interest
to know what this amount would be.

  In a previous paper \cite{ChungYip}, we have already considered the
  finite temperature thermodynamical properties of the nematic
  state, but without the effect of finite magnetization and
  quadratic Zeeman field.   There we in particular have evaluated
  the entropy of the system, and showed that the nematic state can
  tolerate a large entropy without being disordered.
  Since it is now routine that Bose-Einstein condensates be cooled to
  very low temperatures, it should
  therefore be relatively easy to reach this nematic state by ramping
  up an optical lattice from a Bose-Einstein condensate.
   We are therefore particularly hopeful that
  physics of the mentioned spin Hamiltonian can be studied in the
  cold-atom systems.

\begin{figure}
\center
\includegraphics[width=7.0cm]{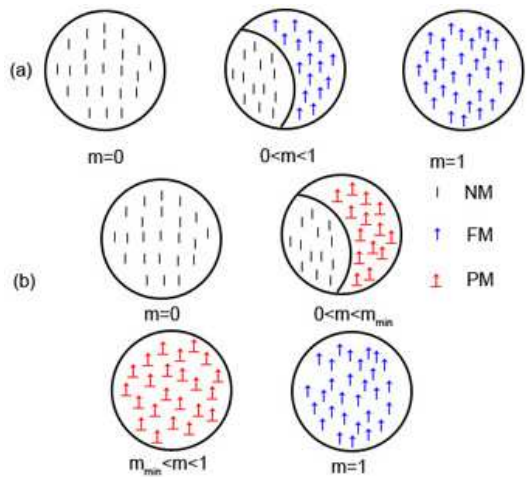}
\caption{(color online) Phase pictures  for zero temperature. $|$ shows the nematic (NM) state, $\uparrow$  the fully magnetic (FM) state and $\underline{\uparrow}$ the partially magnetic (PM) state. (a) For larger $Q$. $m=0$: the NM phase; $0<m<1$: NM and FM phases coexist and are spatially separated. $m=1$: the FM state. (b) For small $Q$.  $m=0$: NM; $0<m<m_{min}$: NM and PM, phase separation; $m_{min}<m< 1$: PM and $m=1$ :FM.
} \label{fig1}
\end{figure}

\begin{figure}
\center
\includegraphics[width=7.0cm]{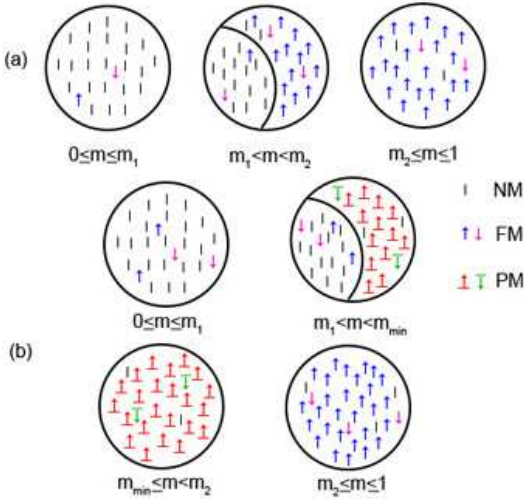}
\caption{(color online) Phase pictures  for finite temperatures. Different symbols are shown. $|$: NM. $\uparrow$: FM with positive magnetization. $\downarrow$: FM with negative magnetization. $\underline{\uparrow}:$ PM with positive magnetization. $\overline{\downarrow}$: PM with negative magnetization. (a) For large $Q$ and (b) for small $Q$. They are quite similar to the zero-temperature case as shown in Fig. \ref{fig1}. However, there still exist some differences. At finite temperature, the system is composed of statistically  mixed states, for example, the NM phase consists of mostly  NM  states  and but mixed with small amounts of FM states.  The PM phase consists of  large amounts of PM states and small amounts of NM states, {\it etc.}. The second difference is that the NM and FM phase have a range of magnetization due to this statistical mixture.}\label{fig2}
\end{figure}

  For the reader's convenience, the  different phases concluded in  this paper  are   pictorially shown in Fig. {\ref{fig1}} for zero temperature and in Fig. {\ref{fig2}} for finite temperatures. At zero temperature, the phases depend on the ground states. For  larger magnitude $Q$ of the quadratic Zeeman effect (we shall provide the condition how large $Q$ should be in the main text) , only two kinds of states appear: the nematic (NM) state with zero magnetization per site $m$ and the fully magnetic (FM) state with $m=1$, as shown in Fig. \ref{fig1}(a). In between the two states coexist and are spatially separated. For smaller magnitude $Q$ the two states remain for $m=0$ (NM)  and $m=1$ (FM), however, a new state appears above the magnetization $m_{min}$: the partially magnetic (PM) state, as shown in Fig. \ref{fig1} (b).  This new state breaks the rotational symmetry along $z$ axis and has magnetization smaller than $1$. If the magnetization is between zero and $m_{min}$, there coexist the NM state and the PM state. For finite temperature, the phase pictures are slightly changed as shown in Fig. {\ref{fig2}}. The system is not a pure state anymore, but a statistical mixture of different states.
For larger $Q$, we have the NM phase if the net magnetization in the system is between $0$ and a small value $m_1$, while the FM phase is obtained if $ m_2 \le m \le 1$.  In between,  phase separation of NM and FM phases is expected. This is shown in Fig. {\ref{fig1}} (a). On the other hand, if $Q$ is small, the PM phase will appear as at zero temperature. The NM phase appears with very small magnetization $m\leq m_1$. The PM phase appears spatially separated from NM above $m_1$ and occupies an increasing volume fraction with increasing magnetization. When $m_{min}$ reached, the PM phase occupies all the region. Above $m_2$, the system is in the FM phase.     

  Our paper is organized as follows. In section \ref{model} the model for a strongly repulsive atom-atom interaction in an optical lattice with linear and quadratic Zeeman effects is introduced. In section \ref{MFT} we provide a mean-field treatment to solve the problem. In section \ref{PD} the phase diagrams are obtained either as a function of magnitude  of  linear Zeeman effect or as a function of the magnetization, both at zero temperature (\ref{PDZeroT}) and at finite temperatures (\ref{PDFiniteT}). In section \ref{Disc} some additional  discussions and the conclusion are made.

\section{Model} \label{model}
In this paper, we consider spin-1 Bosons loaded  in an strong optical lattice under the influence of linear and quadratic Zeeman effects.  In the case of one atom per potential well, such systems can be described by the Hamiltonian
\be \label{Ham}
  H = \sum_{<i,j>} H_{ij} + \sum_i \left( H_i^{L} + H_i^{Q} \right),
\ee
where the two-body Hamiltonian $H_{ij}$ is  related to Bose Hubbard model and $<i,j>$ denotes the next-neighbor sites.  
Defining the hopping constant $t$ and the interaction strength $U_S$ depending on the total spin $S=0,2$,  the on-site repulsion  coefficients in Bose Hubbard model, 
 the energy of the two-body system can be classified according to the total spin and therefore $H_{ij}$ can be written as \cite{Yip03,Imambekov03,ChungYip}  
\be \label{Hij}
  H_{ij} = e_0 P_{ij}^{(0)} + e_2  P_{ij}^{(2)} 
\ee
where $e_0 =-\frac{4t^2}{U_0} $, $e_2 =-\frac{4t^2}{U_2}$ and  the projection operators $P_{ij}^{(S)}$  project the pair $i,j$ into a total spin hyperfine spin $S$ state.
The $H_i^{L}$ term  results from magnetization conservation and the  linear Zeeman splitting\cite{linear}
\be \label{LZ}
   H_i^{L} = - \lambda \left(n_{\uparrow,i} - n_{\downarrow,i}\right)
 \ee
and $H_i^{Q}$ is quadratic Zeeman Hamiltonian
\be
  H_i^{Q}  = 4 Q \left(n_{\uparrow,i}+n_{\downarrow,i} \right)
\ee
with $n_{\uparrow}, n_0$ and $n_{\downarrow}$ representing the number operators with $S_z = 1, 0, -1$, respectively.
The two-body Hamiltonian $H_{ij}$ can also be written in a spin representation \cite{Yip03,Imambekov03}
\be
  H_{ij} =  J({\mathbf S}_i\cdot {\mathbf S}_j) + K ({\mathbf S}_i\cdot {\mathbf S}_j)^2 + J-K, 
\ee 
where  $J = e_2/2, K = (2 e_0+ e_2)/6$ .

\section{Mean-Field Treatment} \label{MFT}

As we have mentioned in our recent paper\cite{ChungYip}, in order to
describe the broken $O(3)$ symmetry for nematic state, 
 one can define a new set of basis, 
\bea
   |x\rangle & = &\frac{1}{\sqrt{2}}(-|\uparrow\rangle + |\downarrow\rangle), \nonumber \\
   |y\rangle & = & \frac{i}{\sqrt{2}}(|\uparrow\rangle + |\downarrow\rangle), \nonumber \\
   |z\rangle & =  &|0\rangle \\.
\eea 
In this basis,  the two-body Hamiltonian $H_{ij}$ can be expressed as a sum of zero order and second order polynomials 
\be \label{HP}
  \begin{split} 
  H_{ij} & = e_0 P_{ij}^{(0)} + e_2 P_{ij}^{(2)} \\
      & = \frac{e_0}{3} \sum_{\alpha.\beta=x,y,z}  |\alpha\rangle_{i} |\alpha\rangle_{j} {_i}\langle\beta|{_j}\langle\beta| + e_2 \sum_{\{I\}} |I\rangle \langle I|,  
  \end{split} 
\ee
where $|I_{\alpha\beta}\rangle = \frac{1}{\sqrt{2}}  (|\alpha\rangle_{i} |\beta\rangle_{j} +  |\beta\rangle_{i} |\alpha\rangle_{j}) $ for $\alpha \neq \beta$, $|I_{0}\rangle = \sqrt{\frac{2}{3}} (|z\rangle_{i} |z\rangle_{j} -\frac{1}{2} |x\rangle_{i} |x\rangle_{j}  -\frac{1}{2} |y\rangle_{i} |y\rangle_{j}) $ and $ |I_{1}\rangle = \frac{1}{\sqrt{2}}  (|x\rangle_{i} |x\rangle_{j} - |y\rangle_{i} |y\rangle_{j})$, and the linear and quadratic Zeeman Hamiltonian have the form
\be \label{eq:LZ}
 H_i^L = -i \lambda \left(| y \rangle_i {_i}\langle x | - | x \rangle_i {_i}\langle y | \right), 
\ee 
\be
 H_i^Q = 4Q \left( | x \rangle_i {_i}\langle x|  + | y \rangle_i {_i}\langle y|  \right). 
\ee
   Without $H^L$ and $H^Q$ terms, we have seen \cite{ChungYip} that the density matrix should have the diagonalized form $\sum_{\alpha=x,y,z} \rho^{\alpha\alpha} |\alpha\rangle \langle \alpha|$. This obviously remains valid when $H^Q$ is included. However, we see that $H^L$ contains off-diagonal terms in the $|x\rangle, |y\rangle$ representation. 
Therefore the general density matrix of a single site should have  the form
\be \label{rho} 
    \hat{\rho} = \sum_{\alpha=x,y,z} \rho^{\alpha\alpha} |\alpha\rangle \langle \alpha| + \rho^{xy} | x \rangle \langle y | + \rho^{yx} | y \rangle \langle x |,  
\ee 
 where $\rho^{\alpha\alpha}$ are real and $(\rho^{xy})^{\star} =  \rho^{yx} $ due to the hermiticity of $\hat{\rho}$. In this way $\rho^{xy}$ and $\rho^{yx}$ can be chosen purely imaginary because together with $\rho^{xx}$ and $\rho^{yy}$ the real part of $\rho^{xy}$ and $\rho^{yx}$ forms a real symmetric matrix and therefore can be diagonalized. In other words, one can rotate the system along $z$-axis to make $\rho^{xy} = -\rho^{yx} = - i\rho^{\|} $ with real number $\rho^{\|}$.    

The principle of mean field theory is to reduce a many-body problem to a one-body problem by replacing all interactions to any one body with an average of effective interaction. A mean-field treatment for a spin-1 Bosons in a lattice has been done by different authors \cite{ChenLevy, ChungYip}. For Hamiltonian (\ref{Ham}), the only term which has to be averaged is the two-body Hamiltonian $H_{ij}$. The effective Hamiltonian to replace $H_{ij}$ be a single site operator
\be
   H_{eff}^0 = z {\mbox{Tr}}_j [ \hat{\rho}_j H_{ij} ]  = z {\mbox{Tr}}_j \left[ \hat{\rho}_j (e_0 P_{ij}^{(0)} + e_2 P_{ij}^{(2)})\right],
\ee
with the coordinate number $z$. For a cubic three-dimensional lattice, $z=6$. Using Eqs.(\ref{HP}) and  (\ref{rho}), $H_{eff}^{0}$ can be obtained as
\be \label{Heff0}
  \begin{split}
  H_{eff}^0 = & z \sum_{\alpha = x,y,z} (K \rho^{\alpha\alpha} + \frac{e_2}{2}) |\alpha\rangle \langle \alpha|\\
              + & z(2J-K) \left\{ \rho^{xy} |x\rangle\langle y| +  \rho^{yx} |y\rangle\langle x|\right\}.
 \end{split} 
\ee
The total  effective Hamiltonian $H_{eff}$ has to include the  linear and quadratic Zeeman effect as well
\be \label{Heff}
  \begin{split}
  H_{eff} = & H_{eff}^0 + H^L + H^Q \\
          = & H_{eff}^0 - i\lambda \left\{|y\rangle \langle x| - |x\rangle \langle y| \right\}  \\ &+ 4 Q   \left\{|x\rangle \langle x| + |y\rangle \langle y|\right\}.  
   \end{split}
\ee
Defining a new set of parameters $h^{\alpha \beta}$ as
\be \label{HParam}
   H_{eff} \equiv -\sum_{\alpha=x,y,z} h^{\alpha \alpha} |\alpha\rangle \langle \alpha| -h^{xy} |x\rangle\langle y| -h^{yx} |y\rangle\langle x|,
\ee
 and comparing Eq. (\ref{HParam})  with Eqs.(\ref{Heff0}) and (\ref{Heff}) we obtain 
\be \label{heq}
  \begin{split}
     h^{xx} & = -z(K\rho^{xx} + \frac{e_2}{2}) -4Q \\
     h^{yy} & = -z(K\rho^{yy} + \frac{e_2}{2}) -4Q \\
     h^{zz} & = -z(K\rho^{zz} + \frac{e_2}{2}) \\
     h^{xy} & = z(K-2J) \rho^{xy} - i\lambda \equiv -i h^{\|}\\
     h^{yx} & = -h^{xy}. 
  \end{split}   
\ee
 where 
\be
  h^{\|} \equiv z(K-2J) \rho^{\|} + \lambda
\ee
and all other components are zero.  
$h^{\alpha\beta}$ can  be one-to-one mapped into $\rho^{\alpha\beta}$ and therefore we can use $h^{\alpha\beta}$ as parameters to find self-consistent equations for the mean-field theory. 

 To find the self-consistent equations we  first rewrite $H_{eff}$ in a matrix representation in $(|x\rangle, |y\rangle, |z\rangle)^{T}$ basis. $H_{eff}$ therefore has the form
\be \label{HSigma}
  H_{eff} = - \left(\sum_{i=0}^3 h_i \sigma_i  + h_z \tau_z \right) 
\ee
where 
\be \label{Sigma}
   \begin{split}
   \sigma_0  &  = \left[ \begin{array}{ccc} 1 & 0 & 0 \\ 0 & 1 & 0 \\ 0& 0& 0 \end{array}\right]   \;\; \sigma_1  = \left[ \begin{array}{ccc} 0 & 1 & 0 \\ 1 & 0 & 0 \\ 0& 0& 0 \end{array}\right]  \;\; \sigma_2 = \left[ \begin{array}{ccc} 0 & -i & 0 \\ i  & 0  & 0 \\ 0& 0& 0 \end{array}\right] \\ 
\sigma_3  &  = \left[ \begin{array}{ccc} 1 & 0 & 0 \\ 0 & -1 & 0 \\ 0& 0& 0 \end{array}\right] \;\; \tau_z  = \left[ \begin{array}{ccc} 0 & 0 & 0 \\ 0 & 0 & 0 \\ 0& 0& 1 \end{array}\right] 
   \end{split}
\ee   
and 
\be \label{h0123}
  \begin{split}
  h_0  = &\frac{h^{xx}+h^{yy}}{2},\;\; h_1 =0, \;\; h_2 = h^{\|},\;\; \\
  h_3  =  &\frac{h^{xx}-h^{yy}}{2},\;\; h_z = h^{zz}.
  \end{split}
 \ee
For the convenience of latter use, we can also define $\rho_i$ and $\rho_z$ in the same way
\be \label{Defrho0123}
  \begin{split}
  \rho_0  = &\frac{\rho^{xx}+\rho^{yy}}{2},\;\; \rho_1 =0, \;\; \rho_2 = \rho^{\|},\;\; \\
 \rho_3  =  &\frac{\rho^{xx}-\rho^{yy}}{2},\;\; \rho_z = \rho^{zz}.
  \end{split}
 \ee
The one-body density matrix of canonical ensemble is defined as 
\be\label{MFrho}
  \hat{\rho} = \frac{\exp{(-\beta H_{eff})}}{{\mbox Tr}\exp{(-\beta H_{eff}})}
\ee
where $\beta \equiv 1/k_B T$.  Inserting Eqs.(\ref{HSigma}), (\ref{Sigma}) and (\ref{h0123}) into Eq. (\ref{MFrho}) and after some algebra (see Appendix A), $\hat{\rho}$ reads
\be \label{rhoSCE}
  \hat{\rho} = \rho_0 \sigma_0 + \rho_2 \sigma_2 + \rho_3 \sigma_3 + \rho_z \tau_z
 \ee
 where
 \be \label{rho0123}
 \begin{split}
 \rho_0 & =  \frac{\cosh{\beta h}}{e^{\beta h_{zo}} + 2 \cosh{\beta h}}, \;\;\; \rho_2   =   \frac{h_2 \sinh{\beta h}}{h(e^{\beta h_{zo}} + 2 \cosh{\beta h})}  \\
  \rho_2  & =  \frac{h_3 \sinh{\beta h}}{h(e^{\beta h_{zo}} + 2 \cosh{\beta h})},  \;\; \rho_z   =  \frac{e^{\beta h_{zo}}}{(e^{\beta h_{zo}} + 2 \cosh{\beta h})} 
 \end{split}
 \ee
with the definitions: $h\equiv \sqrt{h_2^2+h_3^2}$ and $h_{zo} \equiv h_z-h_0$. 

Comparing Eq. (\ref{heq}) with  Eq. (\ref{rhoSCE}), one can obtain three self-consistent equations through the definition of $h_{zo}, h_2$ and $h_3$ in Eq. (\ref{h0123}). The first equation can be obtained by the relation $h_{zo} = -zK(\rho_z-\rho_0)+4Q$, which leads to 
\be \label{hzoSCE}
   h_{zo} = z|K| \left[ \frac{e^{\beta h_{zo}} - \cosh{\beta h}}{e^{\beta h_{zo}} + 2 \cosh{\beta h}}\right] + 4Q.
\ee 
 $h_2 (= z(K-2J)\rho_2+ \lambda)$ accounting for the off-diagonal term in the effective Hamiltonian  gives the second equation
\be \label{h2SCE}
    h_2 = z(K-2J) \frac{h_2 \sinh{\beta h}}{h(e^{\beta h_{zo}}+2\cosh{\beta h})}+\lambda. 
\ee 
  The third equation can be found by the relation: $h_3 = z|K| \rho_3$, which gives the form
\be \label{h3SCE}
   h_3 = h_3 z|K| \frac{\sinh{\beta h}}{h(e^{\beta h_{zo}}+2\cosh{\beta h})}.
\ee
  Therefore there are two situations: if $h_3$ is nonzero, then Eq. (\ref{h3SCE}) can be reduced to 
\be \label{hSCE}
   \frac{h}{z|K|} = \frac{\sinh{\beta h}}{(e^{\beta h_{zo}}+2\cosh{\beta h})}. 
\ee
  Inserting Eq. (\ref{hSCE}) into Eq. (\ref{h2SCE}), $h_2$ is a constant
\be \label{h2}
   h_2 = \frac{|K| \lambda}{2\Delta} 
\ee
with the definition: $\Delta = J-K >0$. In the case that $h_3 =0$, $h = h_2$  and Eq.(\ref{h2SCE}) is also reduced to a two parameter equation
\be \label{h2SCEh30}
   h_2 = z(K-2J) \frac{h_2 \sinh{\beta h_2}}{h(e^{\beta h_{zo}}+2\cosh{\beta h_2})}+\lambda. 
\ee
In either case we have reduced the mean-field problem to two self-consistent equations.  

These self-consistent equations may have many solutions, however, only the one which has the lowest free energy describes the equilibrium state of the system. Therefore we should find the free energy. The free energy can be calculated by the relation
\be \label{FE}
   F  = E_{int} + E_{ext} -TS 
\ee
where the internal energy is given by the two-body interactions
 $
E_{int}=\frac{1}{2} \tr \hat{\rho} H^0_{eff} 
 $
and the external energy is given by the Zeeman fields
 $
E_{ext} = \tr \hat{\rho} \left\{ H^L + H^Q \right\}.
$
After some  algebra (see Appendix B) the free energy is obtained as follows:
\be \label{Eint}
  \begin{split}
   E_{int} = & \frac{zK}{2} \left[ 1 - \frac{4e^{\beta h_{zo}} \cosh{\beta h} + 2}{(e^{\beta h_{zo}} + 2 \cosh{\beta h})^2}\right] \\ & + 2 z \Delta \frac{h_2^2 \sinh^2{\beta h}}{h^2 (e^{\beta h_{zo}} + 2 \cosh{\beta h})^2} + \frac{ze_2}{4}, 
  \end{split}
\ee 
\be \label{Eext}
   \begin{split}
     E_{ext} = & -4Q \frac{e^{\beta h_{zo}}}{e^{\beta h_{zo}} + 2 \cosh{\beta h}} \\
    & - 2\lambda \frac{h_2 \sinh{\beta h}}{h(e^{\beta h_{zo}} + 2 \cosh{\beta h})} +  4Q
   \end{split}
\ee
and 
\be \label{mTS}
  \begin{split}
   -TS = &    \frac{h_{zo} e^{\beta h_{zo}}}{e^{\beta h_{zo}} + 2 \cosh{\beta h}} + \frac{ 2 h \sinh{\beta h}}{e^{\beta h_{zo}} + 2 \cosh{\beta h}}\\
      & - \frac{1}{\beta} \ln{(e^{\beta h_{zo}} + 2 \cosh{\beta h})}. 
  \end{split}
\ee

\section{Phase Diagrams} \label{PD}

In the following, we discuss different phases at zero temperature and at finite temperatures.

\subsection{Zero Temperature} \label{PDZeroT}

\begin{figure}
\center
\includegraphics[width=7.0cm]{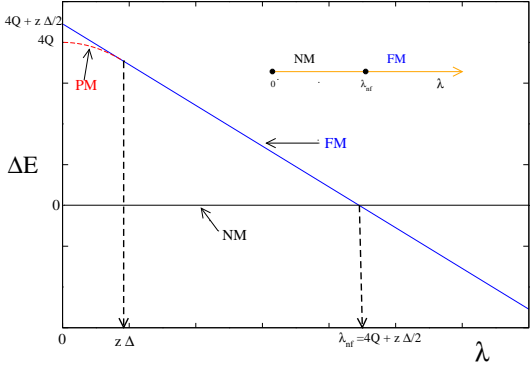}
\caption{(color online)Energy difference of the NM, FM and PM states  with respect to the NM state vs. $\lambda$ for $4Q > z\Delta/2$. The black line shows the reference state: NM, the blue line shows the FM state and red-dashed line shows the PM state. $\lambda_{n,f}$ is the point of a first order phase transition. In the inset we show the phase diagram in terms of $\lambda$. } \label{fig3}
\end{figure} 

At zero temperature, $\rho_{eff}$ is dominated by the smallest eigenvalue of  $H_{eff}$, which can be easily found by diagonalizing Eq. (\ref{Heff}). $H_{eff}$ can be rewritten as follows
\be 
   H_{eff} = - h_0 \mathbbm{1} -h_2 \sigma_2 -h_3 \sigma_3 - h_{zo} \tau_z
\ee
with identity matrix $\mathbbm{1}$. Obviously $h_0  \mathbbm{1} $ is a constant matrix, therefore it can be ignored. We define a new Hamiltonian
\be \label{Heffnew}
   H_{eff}^n = -h_2 \sigma_2 -h_3 \sigma_3 - h_{zo} \tau_z.
\ee

One has three eigenvalues for $H_{eff}^n$: 
eigenvalue $ - h_{zo}$ correspond to eigenvector $[0,0,1]^T(|z \rangle)$ 
and eigenvalues $ \mp h $ correspond to eigenvectors: $[u_{\pm} , v_{\pm},0]^T$, i.e. $(u_{\pm} |x\rangle + v_{\pm} |y \rangle)$.  $u_{-} |x\rangle + v_{-} |y \rangle$ is irrelevant at $T=0 $ because its eigenvalue $h$ is positive. Therefore if $h_{zo} > h$, the system is in the  pure nematic state $|z \rangle$, otherwise the system is in the $(u_{+} |x\rangle + v_{+} |y \rangle)$ state. This state can be either a patially magnetic (PM)  state   or a fully magnetic (FM) state   depending on the parameters $\lambda$ and $q$. We will discuss the details later.  

In the case  $h_{zo} > h$, the free energy can be calculated by using Eqs. (\ref{FE}) - (\ref{mTS}). We can see that at zero temperature the equations show the competition between $h_{zo}$ and $h$. After some algebra, $F$ can be rewritten as a function of $\cosh{\beta h}/e^{h_{zo}}$ and of $\sinh{\beta h}/e^{h_{zo}}$. These two terms disappear at zero temperature. Therefore 
\be
   F = \frac{zK}{2} + \frac{z e_2}{4} \equiv E_z. 
\ee
 We see that $E_z = \frac{zK}{2} + \frac{z e_2}{4} = z (e_0 + 2e_2)/6$ is independent of $\lambda$ and  $Q$.

On the contrary, if $h > h_{zo}$,  the eigenvector of $H_{eff}^n$ can be solved by the equation below
\be
  (h_2 \sigma_2 + h_3 \sigma_3) (u_+ |x \rangle + v_+ |y \rangle) = h (u_+ |x \rangle + v_+ |y \rangle).  
\ee 
 This yields
\be \label{uv}
  u_+ = \frac{1}{\sqrt{2}} (1+\frac{h_3}{h})^{\frac{1}{2}}, \;\;\;
  v_+ = \frac{i h_2}{ \sqrt{2} h} \frac{1}{(1+\frac{h_3}{h})^{\frac{1}{2}}}.   
\ee
In a similar way, by using Eq. (\ref{FE}) - (\ref{mTS}), the free energy defined as $E_+$ has the form 
\be
   F = \frac{z K}{2} + \frac{z e_2}{4} + \frac{z\Delta}{2} \left(\frac{h_2}{h}\right)^2 -\lambda \frac{h_2}{h} + 4Q \equiv E_+. 
\ee
  We can define $h_2/h = \cos\theta$ and $h_3/h =\sin\theta$ since $h^2= h_2^2 + h_3^2$. $E_+$ then takes the form
\be
  E_+ = \frac{z K}{2} + \frac{z e_2}{4} + \frac{z\Delta}{2} \cos^2{\theta} -\lambda \cos{\theta} + 4Q. 
\ee
There are two minima for $E_+$: either
\be \label{theta0}
  \sin{\theta_0} = 0
\ee
or 
\be \label{theta1}
\cos{\theta_1} = \frac{\lambda}{z\Delta}. 
\ee
The second solution has a constraint: $\lambda < z\Delta$, otherwise there is no solution due to the fact that $\cos{\theta}$ can not be larger than $1$.  
These two saddle points can be also obtained by the self-consistent equations. 
In the case  $h_3 =0$, this indicates directly that $\sin{\theta} =0$. This yields $\theta_0 =0$ and then $u = \frac{1}{\sqrt{2}}$ and $v = \frac{i}{\sqrt{2}}$ according to Eq. (\ref{uv}).  Therefore the ground state reads
\be
   |\Psi \rangle = \frac{1}{\sqrt{2}} (|x \rangle + i|y \rangle) = - |\uparrow \rangle. 
\ee 
Therefore we obtain a fully magnetic (FM)  state . 
On the contrary, if $h_3 \neq 0$ at $T=0$, Eq. (\ref{hSCE}) is reduced to the form
\be
  h = \frac{z|K|}{2}. 
\ee
Together with Eq. (\ref{h2}), we obtain Eq. (\ref{theta1}). 
The eigenstate of this solution is 
\be
   |\Psi \rangle = u |x\rangle + v |y\rangle 
\ee
where $u,v(\neq \pm 1) $ are given by Eq.(\ref{uv}). Transforming the state into spin basis, we obtain a state :
\be \label{agstate}
   |\Psi \rangle =  \alpha |\uparrow \rangle + \gamma |\downarrow \rangle
\ee
where
 \be \label{alpha}\alpha = \frac{-u+iv}{\sqrt{2}}  = - \frac{1}{\sqrt{2}}\sqrt{1+ \frac{h_2}{h}} \neq 0\ee and 

\be \label{gamma}\gamma = \frac{u+iv}{\sqrt{2}}   = \frac{1}{\sqrt{2}}\sqrt{1- \frac{h_2}{h}} \neq 0\ee
by using Eq.(\ref{uv}).  
 We call this a partially magnetic (PM)  state. We note that $h_3 \neq 0$ implies that $x$ and $y$ axes are no longer equivalent and the rotational symmetry about the $z$-axis is spontaneously broken in this PM state. 
 
\begin{figure}
\center
\includegraphics[width=7.0cm]{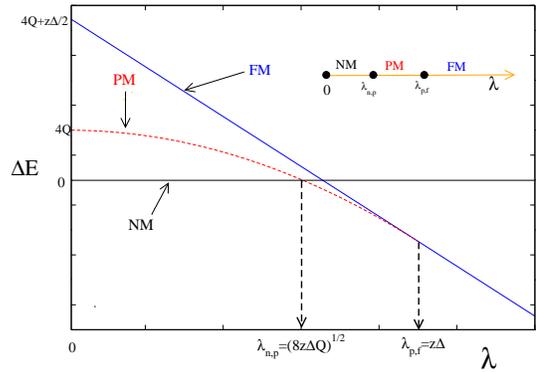}
\caption{(color online) Energy difference of the NM, FM and PM states  with respect to the NM state vs. $\lambda$ for $0< 4 Q<  z\Delta/2$. The black line shows the reference state: NM, the blue line shows the FM state and red-dashed line shows the PM state. $\lambda_{n,p}$ is the point that a  first order phase transition occurs from NM to PM and $\lambda_{p,f}$ is the second-order phase-transition point from PM to FM . In the inset we show the phase diagram in terms of $\lambda$. } \label{fig4}
\end{figure}

We can summarize that we have three phases: nematic state (NM)  $|z \rangle$, FM state $|\uparrow \rangle$ and PM state $\alpha |\uparrow \rangle + \gamma |\downarrow \rangle$. To see which state is preferred we have to calculate the free energy for these three states. 
Define the energy difference first : $\Delta E \equiv E_+ -E_z$. This yields
\be 
 \Delta E(\theta) =  \frac{z\Delta}{2} (\cos{\theta})^2 -\lambda \cos{\theta} + 4Q.
\ee
Therefore a FM state has the energy difference to a NM state
\be \label{E0}
   \Delta E(\theta_0) = \frac{z\Delta}{2} + 4Q - \lambda,
\ee
 while a PM state has the energy difference
\be \label{E1}
  \Delta E(\theta_1) = 4Q - \frac{\lambda^2}{2z\Delta}
\ee  
 with the constraint: $\lambda < z\Delta$. A  state is favored over the  NM states  only if $\Delta E <0$. Therefore a PM state can be a ground state if there exists a critical lambda $\lambda_{n,p}$
\be \label{nag}
  \lambda_{n,p} = \sqrt{8 Q z\Delta} 
\ee
 where $\lambda_{n,p} < z \Delta $.That means  a PM state can be a ground state only with the condition
\be
   4Q < \frac{z\Delta}{2}. 
\ee   
This separates the whole parameter space into two regimes: a regime with PM states and a regime without. 

(a) $4Q \geq \frac{z\Delta}{2}$:
   In this regime, there exist only two states: NM and FM. Fig.~\ref{fig3} shows $\Delta E$ of different states. Since we subtract the energy of the nematic state in the definition of $\Delta E$, we can define $\Delta E =0$ for the nematic state, as the black line shown in Fig.~\ref{fig3}. The blue line decreasing linearly shows $\Delta E(\theta_0)$, the energy difference for FM (\ref{E0}). $\Delta E(\theta_0)$ becomes negative if $\lambda < \lambda_{n,f}$, where
\be
  \lambda_{n,f} = 4Q + \frac{z\Delta}{2}. 
\ee  
The system undergoes a first-order phase transition from nematic states to fully magnetic states while $\lambda$ passing $\lambda_{nf}$.  This picture is also drawn in Fig. {\ref{fig3}}. The reason why the phase transition is first-order is that the  magnetization jumps from zero for nematic states  to one  for ferromagnetic states.  
We note that from Eq.(\ref{LZ}) $\partial \Delta E /\partial \lambda = -(n_{\uparrow}-n_{\downarrow})$, hence the slope of $\Delta E$ versus $\lambda$ is proportional to the magnetization. 
In order to see that the PM state does not appear in this regime, we also draw $\Delta E(\theta_1)$ as the red line in Fig.~{\ref{fig3}}. $\Delta E(\theta_1)$ is always positive till the  end point $\lambda = z\Delta$. Therefore PM never appears in this regime. 

\begin{figure}
\center
\includegraphics[width=7.0cm]{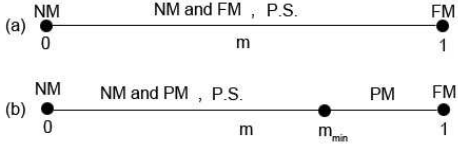}
\caption{(color online) Phase diagrams for zero Temperature in terms of average net magnetization $m$. (a)  $4Q > z\Delta/2$ (b)$0< 4 Q<  z\Delta/2$. P.S. means phase separation. This figure corresponds to Fig. {\ref{fig1}}.} \label{fig5}
\end{figure}

In experiments the magnetization is constant in time,  therefore it is important to have a phase diagram with magnetization as a parameter. Supposed that average magnetization per site is   $m$, the system is purely NM only if $m=0$, while it is purely FM only if $m=1$. In between we have phase separation since the phase transition is first-order. If  $x$ is the fraction of nematic state, then 
 $ x = 1-m$. 
This phase diagram is drawn in Fig.\ref{fig5} (a). 

(b) $0< 4Q < \frac{z\Delta}{2}$:
Fig~{\ref{fig4}} shows $\Delta E$ for different states. $\Delta E(\theta_1)$ is shown with a red dashed line, while $\Delta E(\theta_0)$ with a blue line  as in Fig.~{\ref{fig3}}. We can see that the red dashed line crosses zero at $\lambda_{n,p}$ defined as (\ref{nag}) and then merges to the blue line at the point 
\be
  \lambda_{p, f} \equiv z\Delta. 
\ee
In the regime: $ 0< \lambda < \lambda_{n,p}$ the ground state is nematic, for $  \lambda_{n,p}< \lambda < \lambda_{p,f}$ the system is partially magnetic and one has a fully magnetic  state if $\lambda > \lambda_{p,f}$.  
 Therefore the system undergoes two phase transitions: a first-order phase transition from NM to PM at $\lambda_{n,p}$ and a second-order phase transition from PM to FM at $\lambda_{p, f}$. The second phase transition is second order due to the fact that $\theta_1$ goes to zero while $\lambda$ approaching $z\Delta$, and therefore the transition is continuous for the order parameter.  This yields the phase diagram in the inset of Fig.~{\ref{fig4}}.

The same question arises: if we have a net averaged magnetization per site $m$, which state we will achieve. To see this, we have to calculate the net magnetization for the PM state.  From  Eq.(\ref{agstate}), (\ref{alpha}) and (\ref{gamma})  we can calculate $m$
\be
    m = |\alpha|^2 -|\gamma|^2 = \frac{h_2}{h}, 
\ee
 which leads to 
\be 
   m = \frac{h_2}{h} = \cos\theta_1 = \frac{\lambda}{z\Delta}
\ee
by using Eq. (\ref{theta1}). Therefore in the  regime of PM states, $\lambda_{n,p}< \lambda < \lambda_{p,f}$, the magnetization  lies in the region
\be
  \sqrt{\frac{8Q}{z\Delta}} < m < 1. 
\ee 
 The PM state has a minimum magnetization
\be
     m_{min} =  \sqrt{\frac{8Q}{z\Delta}}. 
\ee
As a result, if $m=0$, the system is purely nematic. For $0< m < m_{min}$ phase separation occurs. One has the nematic state and the PM state spatially separated. Supposed that the fraction in nematic state is defined as $x$, we obtain
$
    x = 1- m\sqrt{\frac{z\Delta}{8Q}} . 
$
 In the regime: $m_{min} < m < 1$,  PM covers the entire system and there exists no nematic state. Finally, if $m = 1$, we obtain FM  again. These results are shown in Fig.\ref{fig5} (b). 

We remark here that the phase diagrams in the inset of Fig. {\ref{fig4}} and Fig. {\ref{fig5}} are analogous to the superfluid case given in Ref.\cite{MIT}. In mean field theories both the lattice and superfluid cases yield mean field energy of the same forms due to symmetry.

\subsection{Finite Temperature} \label{PDFiniteT}

Before we determine the phase diagram for finite temperature, we first figure out different phases by investigating eigenstates of the density matrix $\hat{\rho}$  (\ref{rho}). After diagonalizing it, $\hat{\rho}$ is in its diagonalized form
\be
   \hat{\rho} = P_+ |\Psi_+ \rangle \langle \Psi_+| +  P_- |\Psi_- \rangle \langle \Psi_-|  + P_z |z \rangle \langle z|, 
\ee    
where 
\be \label{Ppm}
   P_{\pm} = \rho_0 \pm \sqrt{\rho_2^2 + \rho_3^2},  \;\;\;\;\; P_z = \rho_z.
\ee  
The eigenvectors read
\be \label{Psi}
    |\Psi_\pm \rangle = u_\pm |x\rangle + v_\pm |y\rangle, 
\ee
where
\be \label{uvpm}
   \begin{split}        
      u_+ & = v_- =  \frac{D}{\sqrt{D^2+\rho_2^2}} \\
      u_- & = v_+ =  \frac{i\rho_2}{\sqrt{D^2+\rho_2^2}}
    \end{split} 
\ee
with the definition: 
\be \label{D}
 D = \rho_3 + \sqrt{\rho_2^2+ \rho_3^2}. 
\ee

\begin{figure}
\center
\includegraphics[width=7.0cm]{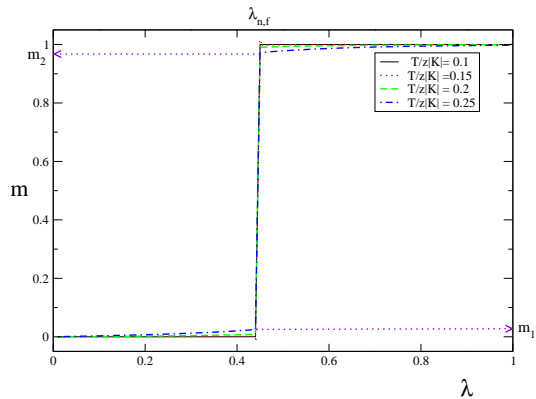}
\caption{(color online)The average magnetization per site $m$ vs. $\lambda$ for different temperatures The parameters: $Q/z|K| = 0.1$, $J/K=0.91$, $\Delta = 0.1 |K|$. We can see that there is a huge magnetization jump at $\lambda_{nf} (\simeq 0.446)$ even for $T/z|K| = 0.25$. The maximum  magnetization in NM is $m_1$ and the minimum magnetization in FM is $m_2$.    } \label{fig6}
\end{figure}

In order to obtain the true phases we have to solve the self-consistent equations  Eqs. (\ref{hzoSCE}) to Eqs. (\ref{h2SCEh30}).  As discussed in the last section, one  can categorize these self-consistent equations into two groups: (1) $h_3 = 0$ and (2) $h_3 \neq 0$ with a constant $h_2$. In the case  $h_3 =0$, $\rho_{xx}=\rho_{yy}$(i.e.$h_3=0$), the eigenvalues (\ref{Ppm}) reads
 \be
   P_{\pm} = \rho_0 \pm \rho_2 = \rho_{xx} \pm |\rho_{xy}|. 
\ee 
According to Eq. (\ref{uvpm}), 
$u_+ = v_- = \frac{1}{\sqrt{2}}$ and $u_- = v_+ = \frac{i}{\sqrt{2}}$,     
 $|\Psi_+\rangle$ thus has the form
\be 
   |\Psi_+ \rangle = \frac{1}{\sqrt{2}} \left(|x\rangle + i |y \rangle \right) = 
    -| \uparrow \rangle, 
\ee
 while $|\Psi_-\rangle$ reads 
\be
   |\Psi_- \rangle = \frac{i}{\sqrt{2}} \left(|x\rangle - i |y \rangle \right) = 
    i | \downarrow \rangle.   
\ee 
Therefore the system is a mixed  state of  $|\uparrow \rangle, |\downarrow \rangle$ and $|z \rangle$.

In the second case that $h_3 \neq 0$ and $h_2 = const.$, the eigenstate $|\Psi_+\rangle$ can be rewritten as 
\be
  |\Psi_+ \rangle = \alpha | \uparrow \rangle + \gamma |\downarrow \rangle    
\ee  
where
 \be
    \alpha = -\frac{1}{\sqrt{2}} (u_+ - i v_+),\;\; \gamma =  \frac{1}{\sqrt{2}}(u_+ + iv_+). 
 \ee
 $\alpha, \gamma$ are nonzero real numbers and $\alpha \neq \gamma $. Similarly, $|\Psi_-\rangle$ has the form
\be
  |\Psi_- \rangle = -\gamma | \uparrow \rangle + \alpha |\downarrow \rangle. 
\ee 
We can easily prove that $|\Psi_+\rangle$ is orthonormal to  $|\Psi_-\rangle$. The system is a mixed state with $(\alpha |\uparrow \rangle + \gamma |\downarrow \rangle), (-\gamma | \uparrow \rangle + \alpha |\downarrow \rangle) $  and $|z \rangle$. 
 
Numerically we solved the self-consistent equations and calculated their free energy according to Eqs. (\ref{FE}), (\ref{Eint}), (\ref{Eext}) and (\ref{mTS}). $J/K = 0.91$ has been used to be close to those for $^{23}Na$. In this case $\Delta = 0.091 |K|$.  In order to find the convergent solution quickly, we start with low temperature ($T/z|K| = 0.1$) and extends the temperature step by step by using the final results as an initial input for the next temperature.
  We have evaluated the phase diagram up to $T/z|K| = 0.25$. We note  that if $\lambda = Q = 0$, the nematic state becomes disordered at $T/z|K| \simeq 0.36$.  
 We illustrate our result with two $Q$ values: $Q/z|K|=0.1$ and $Q/z|K|=0.005$ to represent two regimes as  for Fig. {\ref{fig3}} and Fig. {\ref{fig4}}.   We separate the two sets of self-consistent solutions: one with $h_3=0$ and one with $h_3 \neq 0$ and $h_2 = const.$. For the set of zero $h_3$, two subsets occur. The first one contains the points with small $h_2$: $h_2 \ll 1$ and $h_2$ of the second subset is of order $1$.  Compared with the solutions of zero temperature, the first subset is a continuous evolution with the temperature $T$ from the nematic solution, therefore we can still call these solution nematic (NM),  while in of case of large $h_2$  the solutions correspond to the fully magnetic states (FM) at zero temperature. On the other hand, if $h_3 \neq 0$ and $h_2 =const.$, the states we obtain evolve from the PM states at zero temperature, we can still call them partially magnetic.

\begin{figure}
\center
\includegraphics[width=7.0cm]{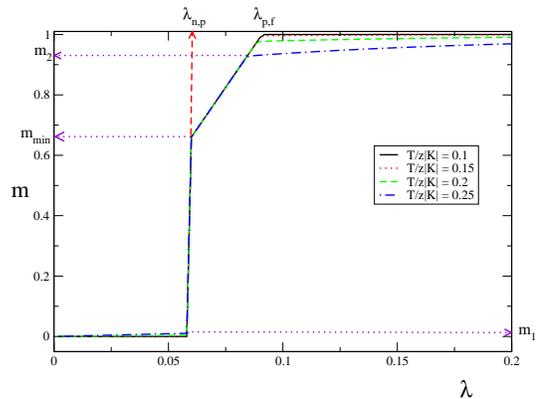}
\caption{(color online) The average magnetization per site $m$ vs. $\lambda$ for different temperatures. $Q/z|K| = 0.005$ and other parameters are same as Fig. {\ref{fig6}}. $m_{min}$ is the minimum magnetization at which PM exists. A first-order phase transition occurs  at $\lambda_{n,p}\simeq 0.06$ from NM to PM and a second-order phase transition occurs at $\lambda_{p,f}$ depending on the temperature.} \label{fig7}
\end{figure}

(a) $Q/z|K| = 0.1$ : in this case, we can calculate the free energy vs. $\lambda$ for the three different sets discussed above. The result is very similar to Fig. \ref{fig3} for each temperature except that the free energy for nematic phase is not constant anymore but a monotonic decreasing function  of $\lambda$. The transition points $\lambda_{n,f}$ of the first order phase transition stay almost the same for all temperatures, for this $Q$, $\lambda_{n,f} \simeq 0.446$. 
The nematic phase at finite temperatures is not a pure state anymore, it contains mostly the  nematic state $|z\rangle$  and with small amounts of $|\uparrow \rangle$ and $|\downarrow \rangle$ due to the fact that $0 \neq h_2 \ll 1$. Therefore the magnetization is not zero. On the other hand, the FM phase is a mixture of large amount of $|\uparrow \rangle$  and small amounts of $|\downarrow \rangle$ and  $|z\rangle$, as a result, the magnetization is smaller than one.  
PM phase can not appear here. 

\begin{figure}
\center
\includegraphics[width=7.0cm]{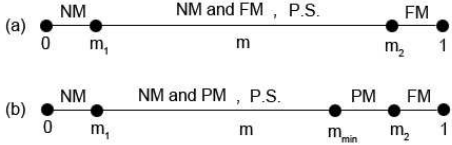}
\caption{Phase diagrams for finite temperatures in terms of average net magnetization $m$. (a)   large $Q$ (b) small $Q$. P.S. means phase separation. This figure corresponds to Fig. {\ref{fig2}}. }\label{fig8}
\end{figure}

We can also calculate magnetization for NM and FM phases by the relation
\be
   m = \tr ( n_{\uparrow} - n_{\downarrow}) \hat{\rho}.
\ee   
It yields
$
   m = 2\rho_2.
$
 Fig. \ref{fig6} shows the magnetization at different temperatures in terms of $\lambda$. $m$  increases monotonically till it reaches its maximum $m_1$ at $\lambda_{n,f}$ and then jumps to value $m_2$. At the end it increases to the fully magnetic state $m=1$ if $\lambda \gg 1$. At higher temperatures, $m_1$ increases and $m_2$ decreases due to the fact that NM and FM mix more and more different states. As a result, if $0\leq m \leq m_1$, the system can be a uniform NM phase, while $m_2 \leq m \leq 1$, we obtain a uniform FM phase. In between, $m_1 < m < m_2$, NM and FM coexist and they are phase separated since the phase transition is first-order. We summarize the result in Fig. {\ref{fig8}} (a).       

 (b) $Q/z|K| = 0.005$: similar to the zero temperature case, PM phase appears here. The  free energy curves vs. $\lambda$ for different temperatures are similar to Fig. {\ref{fig4}}. The first-order phase transition point stays the same: $\lambda_{np} \simeq 0.06$ for all temperatures, while  the second-order phase-transition point $\lambda_{pf}$ changes: $\lambda_{pf} = 0.091, 0.09, 0.088, 0.084$ for $T/z|K| = 0.1, 0.15, 0.2, 0.25$, respectively. The reason is that for PM phase the constraint $h_2 < h$ has to be satisfied, it demands
\be
   \lambda \leq \ z\Delta \frac{2\sinh{\beta h}}{e^{\beta h_{zo}}+ 2\cosh{\beta h}} \equiv h_{p,f}(T). 
\ee
At zero temperature,  $\lambda_{p,f}(0) = z\Delta$ which agrees with the result we obtained in the last section. With increasing $T$ (decreasing $\beta$), $e^{\beta h_{zo}}$ is getting larger, $\lambda_{pf}$ is thus decreasing.

As discussed above, PM phase has large amount of  $(\alpha | \uparrow \rangle + \gamma |\downarrow \rangle)$ mixed with small amounts of $(-\gamma | \uparrow \rangle + \alpha |\downarrow \rangle)$ and $|z\rangle$. The magnetization of the system for different temperatures as a function of $\lambda$ is shown in Fig. \ref{fig7}. In NM phase, $m$ increases and then jumps to $m_{min}$ at $\lambda_{n,p}$. The system undergoes a first order phase transition. In PM phase $m$ ascends to $m_2$ and the system changes continuously to FM phase. We conclude that if $1\leq m \leq m_1$ a homogeneous NM phase is achievable in experiments. In the case that $m_1 < m < m_{min}$, NM and PM phases coexist but separate spatially. In the regime: $m_{min} < m < m_2$, PM phase with different magnetization  is the only phase in the system. In the end, if $m_2\leq m\leq 1$, we obtain FM phase. The phase diagram is plotted in Fig. \ref{fig8} (b).

\section{Discussion and Conclusion} \label{Disc}

\begin{figure}
\center
\includegraphics[width=7.0cm]{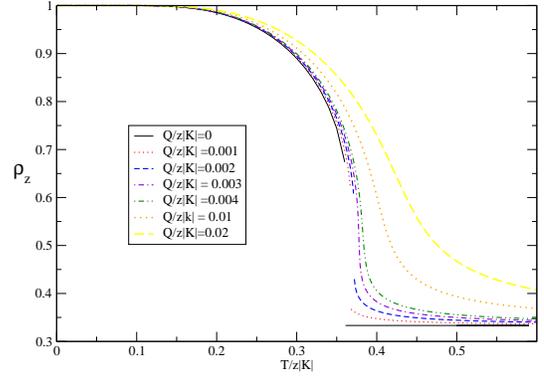}
\caption{(color online)$\rho_z$ vs. temperature for different $Q$'s. In this figure, $\lambda=0$. The first-order phase transition predicted for $Q=0$ still exists till $Q=0.002 z|K|$. Even for larger $Q$, we can still see a rapid decrease of $\rho_z$ when $T$ increases from the low to high temperature regime. } \label{fig9}
\end{figure}

In the case  $\lambda = 0$ and $Q=0$, there exists a first-order phase transition between the nematic state and the disordered state at $T/z|K| = 0.36$ \cite{ChungYip}. The question arises naturally that if $Q$ is nonzero, how the first-order phase transition develops. Fig.~{\ref{fig9}} shows the density $\rho_z$ as a  function of temperature with increasing $Q$. Note that $\rho_z = 1/3$ corresponds to a state with $O(3)$ symmetry.  We found that the first-order phase transition exists till $Q = 0.002 z|K|$ (red-dotted line) and then it turns to be a sharp crossover even till $Q \simeq 0.02 z|K|$ (thick yellow dashed line). From Ref.\cite{MIT}, $Q$ is related to  the magnetic field $B$:
$4Q  = 278*B^2 (Hz G^{-2})$, that means for $Q = 0.002 z|K|$, $B= 0.0134 \sqrt{|K|} G$
with $Hz$ as the unit of $|K|$ . If the superexchange parameter is  $|K| \simeq 100 Hz$\cite{Bloch08},  this  yields $B = 0.134 G$. For $Q = 0.02 z|K|$, $B= 0.42 G$. Experimentally  one can reach $B < 0.01 G$, therefore the first-order phase transition and the sharp crossover can be observable.
On the other hand, as shown in text, the first-order phase transition from the NM state to FM state for large $Q$ and from the NM state to PM state for smaller $Q$ remain at finite temperature. 
We conclude that these phases would phase separate into different spatial regions. One may ask whether, instead of phase separation, one can have, for example, the ferromagnetic sites appear in the form of linear or planar stripes within the nematic regions. 
We exclude this for the following reason. According to Eq. (\ref{Hij}) and $U_2 \simeq U_0 $ for $^{23}Na$, we obtain $e_0 \simeq e_2 < e_1 =0$, where $e_1$ is the energy for total spin for two atoms equal to $1$. 
Consider two neighboring sites.
From the Clebsch-Gordan coefficients we can write down  $|0,0 \rangle$ as a linear combination of  $|F_{tot}=0\rangle$ and $|F_{tot}=2 \rangle$, and $|1,1\rangle$ only exists in $|F_{tot}=2\rangle$, while $|1,0\rangle$ and $|0,1\rangle$ must involve the high energy  $|F_{tot}=1\rangle$ state. Therefore it costs more energy if the system builds a domain wall than just put the same $m_f$ state as neighbors. That is the reason why the system prefers a spatially separated phase than stripe phases. A stripe phase is not favored   because it needs to build more than one domain wall. 

To conclude, we have shown the phase diagrams for a spin-1 polar Bose gas loaded in an strongly repulsive optical lattice. There exist three different phases: the nematic (NM), fully magnetic (PM) and partially magnetic (PM) phases depending on the parameter regime of  the system. A first-order phase transition from NM to FM or from NM to PM has been predicted. A second-order phase transition from PM to FM is also found. These phase transitions are robust even at finite temperatures. Therefore they should be observable in experiments.

\appendix
\section{ Density Matrix}
To obtain $\hat{\rho}$ defined as Eq. (\ref{MFrho}), we have to calculate $e^{-\beta H_{eff}}$ first. It can be written in the form
\be  \label{expH}
  e^{-\beta H}  = e^{\beta \left(\sum_{i=0}^3 h_i \sigma_i + h_z \tau_z \right)}. 
 \ee
Since $\tau_z$ commutes with all $\sigma_i$, we can take $e^{\beta h_z \sigma_z}$ out of the exponential. One can prove the relation with properties of Pauli matrices: 
\be
  e^{\xi \hat{n} \cdot \vec{\sigma}} = \cosh{\xi} +  \hat{n} \cdot \vec{\sigma} \sinh{\xi}, 
\ee
 where $\hat{n}$ is a three dimensional normal vector and $\vec{\sigma} = [\sigma_1, \sigma_2, \sigma_3]^T$. By using this, Eq.(\ref{expH}) has the form 
\be
   \begin{split}
    e^{-\beta H}  = & e^{\beta h_{z}} \tau_z + e^{\beta h_0} \cosh{\beta h}\; \sigma_0  \\
     & + \left(\frac{h_2}{h} \sigma_2 + \frac{h_3}{h} \sigma_3 \right)  e^{\beta h_0} \sinh{\beta h}. 
  \end{split}
\ee
It yields
 \be
    \tr e^{-\beta H} = e^{\beta h_{z}}  + 2  e^{\beta h_0} \cosh{\beta h}. 
 \ee
  Eq. (\ref{rhoSCE}) together with Eq. (\ref{rho0123}) are thus obtained.

\section{ Free Energy} 

The internal energy can be calculated by using Eq.(\ref{Heff0}) as follows
\be
  \begin{split} 
  E_{int} = & \frac{1}{2} \tr \hat{\rho} H^0_{eff} \\
          = &   \frac{1}{2} z \sum_{\alpha = x,y,z} (K \rho^{\alpha\alpha} + \frac{e_2}{2}) \rho^{\alpha \alpha} \\
            & +   z\left(J-\frac{K}{2}\right) \left\{ \rho^{xy} \rho^{yx} +  \rho^{yx} \rho^{xy}. 
 \right\}
  \end{split}
\ee
 This term can be simplified to a form
\be \label{Eint1}
  E_{int} = \frac{zK}{2} \tr {\hat{\rho}}^2 + 2z (J-K) \rho_2^2 + \frac{ze_2}{4}.  
\ee
Note that $\tr {\hat{\rho}}^2 = 2 \rho_0^2 + 2\rho_2^2 + 2 \rho_3^2 + \rho_z^2 $.Inserting Eq. (\ref{rho0123}) into Eq. (\ref{Eint1}), we obtain internal energy as Eq. (\ref{Eint}).  $E_{ext}$ can be calculated in a similar way:
\be
   \begin{split}
      E_{ext} = & \tr \hat{\rho} \left\{ 4Q (|x\rangle \langle x| + |y \rangle \langle y|) \right. \\  & \left. +  i \lambda (|x \rangle \langle y| -| y\rangle \langle x|) \right\} \\
      = & 4Q- 4Q \rho^{zz} - 2 i\lambda \rho^{xy},  
   \end{split}
\ee 
   this yields
   \be 
     E_{ext} = 4Q -4Q\rho_z -2\lambda \rho_2. 
   \ee
  Therefore Eq. (\ref{Eext})  is obtained by using Eq. (\ref{rho0123}).

Finally $-TS$ can be obtained by the definition of entropy:
\be
   S \equiv - k_B\tr \hat{\rho} \ln \hat{\rho}. 
\ee
It yields
 \be
   \begin{split}
   -TS = & -\tr  \hat{\rho} H_{eff} -\frac{1}{\beta} \ln \tr e^{\beta H_{eff}} \\
        = & (2\rho_0-1) h_0 + 2\rho_2 h_2 + 2 \rho_3 h_3 + \rho_z h_z \\ & -\frac{1}{\beta} \ln{(e^{\beta h_{zo}}+2\cosh{\beta h})}, 
    \end{split}
\ee 
  and therefore Eq.({\ref{mTS}}).

\vspace{4cm}



\begin{thebibliography}{99}
\bibitem{linear} The linear Zeeman field splitting combines with
the Lagrange multiplier for the constant magnetization constraint
 to form the coefficient $\lambda$ in Eq. (\ref{LZ}) and
therefore does not yield an additional physical parameter.
\end{thebibliography}
\end{document}